\documentclass{jpp}

\newcommand{\apj}{\mbox{\it Astrophysical Journal}}

\usepackage{color}
\usepackage{graphicx}
\usepackage{natbib}

\ifCUPmtlplainloaded \else
  \checkfont{eurm10}
  \iffontfound
    \IfFileExists{upmath.sty}
      {\typeout{^^JFound AMS Euler Roman fonts on the system,
                   using the 'upmath' package.^^J}%
       \usepackage{upmath}}
      {\typeout{^^JFound AMS Euler Roman fonts on the system, but you
                   dont seem to have the}%
       \typeout{'upmath' package installed. JPP.cls can take advantage
                 of these fonts, if you use 'upmath' package.^^J}%
      }
  \else
  \fi
\fi

\ifCUPmtlplainloaded \else
  \checkfont{msam10}
  \iffontfound
    \IfFileExists{amssymb.sty}
      {\typeout{^^JFound AMS Symbol fonts on the system, using the
                'amssymb' package.^^J}%
       \usepackage{amssymb}%
         
         \let\geq=\geqslant
      }{}
  \fi
\fi

\ifCUPmtlplainloaded \else
  \IfFileExists{amsbsy.sty}
    {\typeout{^^JFound the 'amsbsy' package on the system, using it.^^J}%
     \usepackage{amsbsy}}
    {}
\fi

\newsavebox{\astrutbox}
\sbox{\astrutbox}{\rule[-5pt]{0pt}{20pt}}

\title[Density jump as a function of field in collisionless shocks]{Density jump as a function of magnetic field strength for parallel collisionless shocks in pair plasmas}

\author[A. Bret and R. Narayan]%
{A\ls N\ls T\ls O\ls I\ls N\ls E\ns B\ls R\ls E\ls T$^{1,2,3}$, R\ls A\ls M\ls E\ls S\ls H\ns N\ls A\ls R\ls A\ls Y\ls A\ls N$^3$%
  \thanks{Email address for correspondence: antoineclaude.bret@uclm.es}
}

\affiliation{$^1$ETSI Industriales, Universidad de Castilla-La Mancha, 13071 Ciudad Real, Spain\\[\affilskip]
$^2$Instituto de Investigaciones Energ\'{e}ticas y Aplicaciones Industriales, Campus Universitario de Ciudad Real, 13071 Ciudad Real, Spain\\[\affilskip]
$^3$Harvard-Smithsonian Center for Astrophysics, Harvard University, 60 Garden St., Cambridge, MA 02138 USA
}

\date{?; revised ?; accepted ?. - To be entered by editorial office}

\begin{document}

\maketitle

\begin{abstract}
Collisionless shocks follow the Rankine-Hugoniot jump conditions to a good approximation. However, for a shock propagating parallel to a magnetic field, magnetohydrodynamics states that the shock properties are independent of the field strength, whereas recent Particle-in-Cell simulations reveal a significant departure from magnetohydrodynamics behavior for such shocks in the collisionless regime. This departure is found to be caused by a field-driven anisotropy in the downstream pressure, but the functional dependence of this anisotropy on the field strength is yet to be determined. Here, we present a non-relativistic model of the plasma evolution through the shock front, allowing for a derivation of the downstream anisotropy in terms of the field strength. Our scenario assumes double adiabatic evolution of a pair plasma through the shock front. As a result, the perpendicular temperature is conserved. If the resulting downstream is firehose stable, then the plasma remains in this state. If unstable, it migrates towards the firehose stability threshold. In both cases, the conservation equations, together with the relevant hypothesis made on the temperature, allows a full determination of the downstream anisotropy in terms of the field strength.
\end{abstract}

\maketitle

\section{Introduction}

When a shockwave propagates along a magnetic field $\mathbf{B}_0$, magnetohydrodynamics (MHD) states that the fluid is disconnected from the field  \citep{Lichnerowicz1976, Majorana1987}.
As a result, the density jump at the front does not depend on $B_0$.
However, recent Particle-in-Cell (PIC) simulations of such parallel relativistic shocks in collisionless pair plasmas found that, with increasing $B_0$, the density jump becomes progressively smaller than the MHD prediction \citep{BretJPP2017}.
 This behaviour could be traced to the inability of the downstream plasma to efficiently isotropize the particle distribution, as the field tends to guide particles and generate an anisotropic distribution function \citep{BretJPP2016}.

 Regardless of whether the system is relativistic or not, one expects the plasma to display a 1D behavior in the limit of infinite field strength. For example, the density jump for a non-relativistic strong shock in this limit should tend to $(\Gamma_{1D}+1)/(\Gamma_{1D}-1)=2$, with the adiabatic index of a 1D gas $\Gamma_{1D}=3$.

Some authors have already dealt with  anisotropic distributions in magnetized shocks, which eventually result in anisotropic pressures \citep{Karimabadi95,Vogl2001}.
In this respect, \cite{Erkaev2000} and \cite{Vogl2001} studied the jump conditions for non-relativistic perpendicular and oblique shocks, while \cite{Gerbig2011}  computed the jump conditions for relativistic MHD shocks in a gyrotropic plasma.

 The following sentences from \cite{Erkaev2000} describe well a common feature of many such studies: ``The ratio of the perpendicular and parallel plasma pressures downstream of the shock is an unknown parameter that has to be determined. In principle, this parameter depends on the structure of the shock; however, this is beyond the scope of an MHD model''. That is, MHD per se cannot predict the anisotropy. Still, PIC simulations like the ones conducted in \cite{BretJPP2017} in the relativistic regime, show a deterministic downstream anisotropy in terms of the initial parameters of the problem.

 The novelty of the present work is that, while we work within an MHD formalism, we do not treat the
 downstream anisotropy as a free parameter, only constrained by some
 instability thresholds. Rather, we compute the anisotropy by assuming a certain
 kinetic history of the plasma through the shock front. In spirit our
 work is somewhat similar to \cite{Vogl2001}, who also solved for the
 anisotropy in the post-shock plasma. Their work focused on the mirror
 instability, which is relevant for their perpendicular and oblique
 shocks, whereas we consider the firehose instability, which is
 relevant for our parallel shocks. In addition, as we explain below,
 we find that it is necessary to consider the problem as a two-stage
 process (Stage 1, Stage 2), but \cite{Vogl2001} ignore this
 complication.

 A conundrum in this respect is the following. While in the absence of a magnetic field the Vlasov equation predicts that anisotropic distributions are unstable \citep{Weibel},
 it does not impose a unique degree of anisotropy in the presence of a field.
 Proofs of this are the beautiful studies of the solar wind, which show that the parallel and perpendicular temperatures, while limited by the thresholds for mirror
 and firehose instabilities, populate all regions of the stable zone \citep{BalePRL2009,MarucaPRL2011,SchlickeiserPRL2011}. As a result, a full range of anisotropy degrees can be found in the solar wind, all of them stable\footnote{The solar wind plasma does migrate towards $T_\perp = T_\parallel$, but under the effect of collisions. See the ``collisional age'' plot in \cite{BalePRL2009}.}.

In the context of shock physics, one might expect the downstream anisotropy to be a function of the field strength and of the upstream properties \citep{BretJPP2017}. A determination of this functional dependence would allow for a modification of MHD codes, making them capable of mimicking the effects of the underlying kinetic dynamics.

\begin{figure}
  \begin{center}
   \includegraphics[width=.8\textwidth]{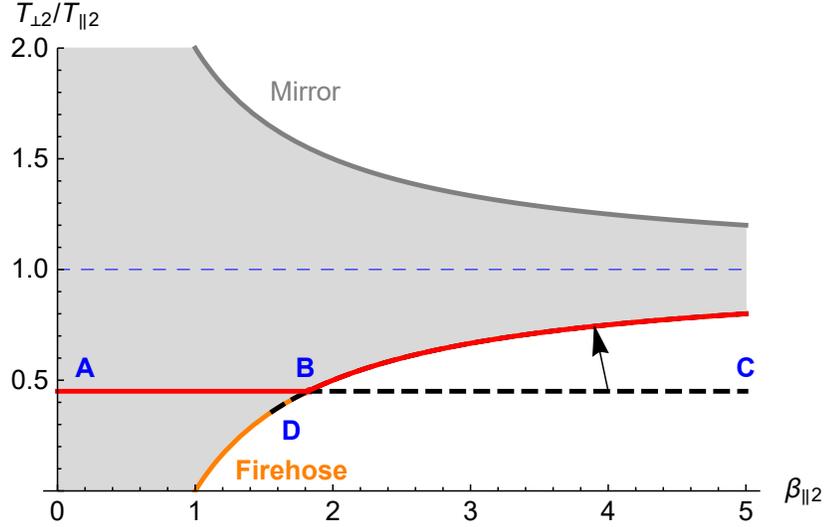}
  \end{center}
\caption{When the plasma arrives downstream (subscripts ``2''), it comes with a certain anisotropy constrained by the condition $T_{\perp 2}=T_{\perp 1}$, and a certain $\beta_{\parallel 2}$ parameter. As explained in the text (see the bullets points after Eq. \ref{eq:firehose}), our scenario implies $T_{\perp 2}/T_{\parallel 2}<1$ (the line ABC for the case considered here). If $\beta_{\parallel 2}$ is such that the plasma lies inside the firehose stable region (gray shaded region, segment AB), then it remains as it is. But if the plasma arrives in an unstable region of the phase space (segment BC), it will move to a point on the firehose threshold (arrow). Thus, the end state of the plasma is somewhere on the red line. The short dashed line on the firehose threshold curve, between \textbf{B} and \textbf{D}, is explained in Section \ref{sec:contact}.}
\label{fig:firehose}
\end{figure}

\bigskip

As a first step towards an understanding of this problem, we present a scenario for the history of the plasma through the shock front. Consider Figure \ref{fig:firehose} which shows the downstream phase space $(\beta_{\parallel 2},~T_{\perp 2}/T_{\parallel 2})$, with
\begin{equation}\label{eq:beta}
\beta_{\parallel 2} = \frac{n_2 k_B T_{\parallel 2}}{B_0^2/8\pi},
\end{equation}
where the subscript ``2'' stands for the downstream properties (``1'' stands for the upstream), ``$\parallel, \perp$'' refer to ``parallel'' and ``perpendicular'' to the field
 respectively, and $k_B$ is the Boltzmann constant. The field $\mathbf{B}_0$ is aligned with the $x$ axis, which is also the direction of the flow, i.e., we consider parallel (non-relativistic) shocks.

 In the region $T_{\perp 2}/T_{\parallel 2}<1$, the stability of the plasma is limited by the firehose instability threshold \citep{Gary1993},
\begin{equation}\label{eq:firehose}
\frac{T_{\perp 2}}{T_{\parallel 2}}=1-\frac{1}{\beta_{\parallel 2}},
\end{equation}
while in the region $T_{\perp 2}/T_{\parallel 2}>1$, it is limited by the mirror instability threshold, $T_{\perp 2}/T_{\parallel 2}=1+1/\beta_{\parallel 2}$. As discussed below, the latter is not of interest for the problem considered in this paper.

The scenario we conjecture for the history of the plasma is illustrated by the flow-chart in Figure \ref{fig:history}. It goes as follow:

\begin{figure}
  \begin{center}
   \includegraphics[width=.9\textwidth]{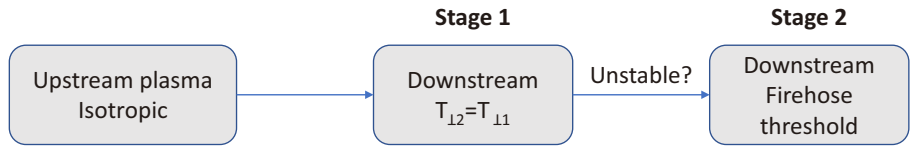}
  \end{center}
\caption{Proposed scenario for the plasma, from the upstream to the downstream. The upstream plasma is isotropic. Then, across the shock front, we assume that $T_\perp$ is conserved. This is ``Stage 1''. If the resulting plasma is stable, then this is the end state. If firehose-unstable, then the plasma exchanges energy between perpendicular and parallel particle motions until the firehose stability threshold is reached. This is ``Stage 2''. In each case, the properties of the plasma are determined by the conservation equations (\ref{eq:conser1}-\ref{eq:conser3}).}
\label{fig:history}
\end{figure}

\bigskip

\begin{itemize}
 \item For simplicity, we assume that the upstream plasma is isotropic.

  \item As the plasma moves from upstream to downstream, we assume
    that its perpendicular temperature is initially conserved.  This
    can be justified at the macroscopic (MHD) level, since $T_\perp
    \propto B_0$ during adiabatic evolution of the plasma. For a
    parallel shock, $B_0$ does not change through the shock front,
    hence $T_\perp$ should not either. In contrast, $T_\parallel$
    should increase as $T_\parallel \propto n^2/B_0^2$, under the same
    adiabatic assumption (\citealt{CGL1956} or
    \citealt{bittencourt2013}, p. 304).

  \item In addition to the above anisotropic heating from
    anisotropic compression, irreversible processes at the shock front
    will generate entropy. We assume that this additional energy again
    goes into parallel motions and causes only $T_\parallel$ to
    increase. As a result of this and the previous assumption, the
    plasma ``lands'' in the downstream with some anisotropy $T_{\perp
      2}/T_{\parallel 2} < 1$, corresponding to the line
    \textbf{A}-\textbf{B}-\textbf{C} in Fig. \ref{fig:firehose}. We call this
    ``Stage 1''.

  \item If the point in the phase space $(\beta_{\parallel 2},
    ~T_{\perp 2}/T_{\parallel 2})$ where the plasma lands at the end of
    Stage 1 is between \textbf{A} and \textbf{B}, the plasma is stable
    and we assume that nothing further happens to the (collisionless)
    plasma. However, if the point is between \textbf{B} and
    \textbf{C}, the plasma is firehose unstable. In this case, we
    assume that the plasma moves towards a state lying on the firehose
    stability threshold (arrow). We call this ``Stage 2''.

    When the plasma at the end of Stage 1 is unstable, the mechanism responsible for the switch to Stage 2 is the firehose instability. The characteristic time of the transition should be of the order of the inverse growth-rate of the instability. The path from stage 1 to stage 2 could be assessed numerically or guessed through quasi-linear theory. However, it is not relevant for our purposes as only the end of the path matters, which is uniquely determined by the conservation equations (\ref{eq:firehose}-\ref{eq:conser3}).
\end{itemize}

\bigskip

The assumptions encoded in bullets 3 and 4 above are rather
strong. The main support comes from the macroscopic double adiabatic
theory of \cite{CGL1956}, coupled with the fact that it
appear to be reasonable at least in the limit of infinite field
strength.

As a result of the double adiabatic evolution of the plasma through the front, the perpendicular temperature is conserved, while the parallel one increases. Since we assumed $T_{\perp 1}/T_{\parallel 1}=1$ for the upstream, the first stage of the downstream history has $T_{\perp 2}/T_{\parallel 2} < 1$. Therefore, and as evidenced on Fig. \ref{fig:firehose}, only the firehose instability is relevant to our scenario, since the mirror instability only limits the opposite $T_{\perp 2}/T_{\parallel 2} > 1$ range. Note that it does not mean that the mirror instability is not important for all collisionless shocks under more general scenarios (see for example \citealt{Vogl2001} and \citealt{Kunz2014}). Indeed, according to our scenario, the mirror instability would be relevant for perpendicular shocks (see conclusion).

The extension of the double adiabatic
theory to a finite field strength
is the key ansatz of the present paper. Although this ansatz should
apply regardless of the plasma composition, we here restrict our
treatment to the case of a pair plasma. The reason for this is that
there is heating at the shock front, due to both compression and
entropy generation. Both processes could affect electrons and ions
differently, resulting in a downstream plasma with different electron
and ion temperatures \citep{Guo2017,Guo2018}. In a pair plasma,
electrons and positrons will undergo identical heating, so we can
speak of a single perpendicular temperature and a single parallel
temperature. This simplifies the problem considerably. Note that the
firehose and mirror instabilities in pair and electron/ion plasmas are
similar \citep{Gary2009,Schlickeiser2010}, so Eq. (\ref{eq:firehose})
and Fig.~\ref{fig:firehose} hold for a pair plasma.

In addition to differential heating at the front, ion/electron collisionless shocks invoke acoustic waves or ion cyclotron resonance for example, as part of the heating mechanism on the downstream. These kinetic effects could play a major role in many astrophysical plasmas of interest. It is therefore important to keep in mind that although pair plasmas are interesting from the theoretical and numerical points of view, translating the current approach to solar system shocks, for instance, may require bridging important gaps.

As a result of the history outlined above, the downstream plasma
eventually settles somewhere on the thick red line in
Fig. \ref{fig:firehose}.  Since $\beta_{\parallel 2} \propto
B_0^{-2}$, we expect that, with increasing $B_0$, the downstream
plasma will move from right to left on this line. In the limit
$B_0=0$, the point representing the plasma in the $(\beta_{\parallel 2}, T_{\perp 2}/T_{\parallel 2})$ phase space is located to the far right of the figure and the plasma
will be highly unstable after Stage 1. The plasma will then
undergo Stage 2 and will end on the firehose stability threshold,
which in this case will correspond to perfect isotropy. For higher
fields $B_0$, the plasma will continue to be firehose unstable,
and will move to a state of marginal stability via Stage 2, though now
the final state will involve some anisotropy. Above some critical
value of $B_0$, Stage 1 will result in the plasma ending up in the
segment \textbf{A}-\textbf{B} in Fig. \ref{fig:firehose}. There will no longer be any need for
Stage 2.

The rest of the article is dedicated to the determination of the
downstream properties for each of the two scenarios described above:
Stage 1, Stage 2. The required equations are simply the non-relativistic conservation equations for matter, momentum and energy,
\begin{eqnarray}
  n_1 V_1                                  &=&  n_2 V_2,         \label{eq:conser1} \\
  n_1 V_1^2 + P_1                          &=&  n_2 V_2^2 + P_{2x}, \label{eq:conser2} \\
  \frac{V_1^2}{2} + \frac{P_1}{n_1} + U_1  &=&  \frac{V_2^2}{2} + U_2 + \frac{P_{2x}}{n_2}, \label{eq:conser3}
\end{eqnarray}
where $U$ is the internal energy and $x$ corresponds to the direction of the flow
as well as the orientation of the magnetic field.

Note that, because the downstream is not isotropic, the downstream pressure entering equation (\ref{eq:conser2}) is the component of pressure parallel to $x$, since this equation balances the momentum gained with the pressure force along the direction of motion. Similarly, Eq. (\ref{eq:conser3}) only accounts  for the $x$ component of $P_2$ because, while $U_i$ is the total internal energy (both parallel and perpendicular), the pressure term, which arises from the work done by the pressure force on the fluid, involves only the $x$ (parallel) component (see for example \cite{FeynmanVol2}, \S 40-3).

For the isotropic upstream, we shall always use $U_1=  (3/2)P_1/n_1$. But for the downstream, the expression for $U_2$ changes according to the anisotropy.

We shall use in the rest of this paper the following dimensionless variables,
\begin{eqnarray}\label{eq:dimless}
  r          &=& \frac{n_2}{n_1}, \nonumber\\
  A_2          &=& \frac{T_{\perp 2}}{T_{\parallel 2}}, \nonumber\\
  \sigma   &=&  \frac{B_0^2/8\pi}{n_1V_1^2/2}, \nonumber\\
  \chi_1^2 &=& \frac{V_1^2}{P_1/n_1}.
\end{eqnarray}

The parameter $r$ denotes the density, or compression, ratio; $A_2$ stands for the downstream anisotropy ratio ($A_1=1$ since the upstream is assumed isotropic); $\sigma$ measures the strength of the magnetic field through the ratio of magnetic to kinetic energy densities.

The parameter $\chi_1$ obviously resembles a Mach number. It is nevertheless convenient to defer its physical interpretation until the analysis of the two Stages is completed.

The natural dimensionless parameter for the field seems to be $\beta_{\parallel 2}$. However, $\sigma$ is commonly used in PIC simulations of collisionless shocks \citep{Sironi2011ApJ,Marco2016}. We therefore conduct the main part of our analysis using $\sigma$. We will come back to $\beta_{\parallel 2}$ in section \ref{sec:betapara}.

In section \ref{sec:mod1}, we solve the conservation equations assuming $T_{\perp 2}=T_{\perp 1}$, thus defining the properties of Stage 1.
 In section \ref{sec:mod2}, we solve the same equations but assuming $T_{\perp 2}/T_{\parallel 2}=1-1/\beta_{\parallel 2}$ instead. This corresponds to Stage 2. Following these two analyses, the physical interpretation of the parameter $\chi_1$ is given in section \ref{sec:X}.   Sections \ref{sec:contact} and \ref{sec:betapara} then explain how the two Stages fit together, eventually providing a coherent picture of the shock properties in terms of the magnetic field strength.

\section{Stage 1: Downstream with $T_{\perp 2}=T_{\perp 1}$}\label{sec:mod1}
We here determine the density jump and the downstream anisotropy under the assumptions of Stage 1, that is, assuming conservation of the perpendicular temperature.

\subsection{Density jump and anisotropy}\label{subsec:stag1}

In general we have $U=(P_x+P_y+P_z)/2n$, which reduces to $U=(3/2)P/n$  for an isotropic plasma. In the present case, the downstream is anisotropic, with a parallel pressure $P_{2x}$  different from the perpendicular pressures $P_{2y}=P_{2z}\equiv P_{\perp 2}$. We can therefore write $U_2=(P_{2x}+2P_{\perp 2})/2n_2$.
Since $P_{\perp 2} = n_2 k_BT_{\perp 2}$, and since we assume $T_{\perp 2}=T_{\perp 1}$, we have
\begin{eqnarray}
  U_2 &=& \frac{1}{2n_2}(P_{2x}+2n_2 k_BT_{\perp 2}) = \frac{1}{2n_2}(P_{2x}+2n_2 k_BT_{\perp 1})\nonumber\\
      &=& \frac{1}{2n_2}\left(P_{2x}+2\frac{n_2}{n_1} P_1\right) \nonumber\\
      &=& \frac{ P_{2x}}{2n_2} + \frac{P_1}{n_1}  .  \label{eq:U2}
\end{eqnarray}
From Eqs. (\ref{eq:conser1}, \ref{eq:conser2}) we obtain,
\begin{equation}\label{eq:Px1}
  P_{2x} =  n_1 V_1^2 + P_1 - n_2 \left( V_1 \frac{n_1}{n_2} \right)^2 .
\end{equation}
Eqs. (\ref{eq:conser3}, \ref{eq:U2}) then give,
\begin{equation}\label{eq:Px2}
  P_{2x} = \frac{2}{3} n_2 \left( \frac{V_1^2}{2}-  \frac{1}{2}\left( V_1 \frac{n_1}{n_2} \right)^2 + \frac{3}{2}\frac{P_1}{n_1} \right)  .
\end{equation}
Equating Eqs. (\ref{eq:Px1}) and (\ref{eq:Px2}) gives a polynomial equation for $n_2$. One root is the trivial solution $n_2=n_1$. The other root is,
\begin{equation}\label{eq:n2}
  n_2 =  n_1 \frac{2 n_1 V_1^2}{3P_1 + n_1 V_1^2} .
\end{equation}
Inserting this root into Eq. (\ref{eq:Px1}) then gives,
\begin{equation}\label{eq:PxOK}
  P_{2x} =  \frac{1}{2}(n_1 V_1^2-P_1) = n_2 k_BT_{ \parallel 2},
\end{equation}
so that with (\ref{eq:n2}),
\begin{equation}\label{eq:Tperp2}
  k_BT_{\parallel 2} = \frac{P_{2x}}{n_2} =  \frac{\left(n_1 V_1^2-P_1\right) \left(n_1 V_1^2+3 P_1\right)}{4 n_1^2 V_1^2}.
\end{equation}
Regarding the anisotropy ratio in the downstream, we now use $k_BT_{\perp 2}=k_BT_{\perp 1}=P_1/n_1$, so that
\begin{equation}\label{eq:AniRatio}
  \frac{T_{\perp 2}}{T_{\parallel 2}} = \frac{4 n_1 P_1 V_1^2}{\left(n_1 V_1^2-P_1\right) \left(n_1 V_1^2+3 P_1\right)}.
\end{equation}
Introducing the dimensionless variables defined in (\ref{eq:dimless}), we  obtain for (\ref{eq:n2}) and (\ref{eq:AniRatio}),
\begin{eqnarray}\label{eq:AniRatioMachOK}
r  &=& \frac{2 \chi_1^2 }{\chi_1^2+3}, \nonumber \\
A_2  &=& \frac{4 \chi_1^2}{\chi_1^4 + 2 \chi_1^2-3} .
\end{eqnarray}

The above result shows that the anisotropy $A_2$ goes to 0 in the limit $\chi_1 \gg 1$. In the same limit, the density ratio $r$ goes to 2. As mentioned earlier, this is the expected density jump for a strong non-relativistic 1D shock.
This result will be further discussed in section \ref{sec:X}.

\subsection{Stability}
The downstream plasma can fulfill the conditions for Stage 1, namely, $T_{\perp 2}=T_{\perp 1}$, and will remain there so long as it is stable. It is therefore relevant to assess the firehose-stability of the plasma. To do so, one needs to check whether it satisfies the firehose stability criterion (\ref{eq:firehose}).

The parameter $\beta_{\parallel 2}$ is given by,
\begin{equation}\label{eq:BetaMod1}
\beta_{\parallel 2} = \frac{2r}{\sigma A_2 \chi_1^2}.
\end{equation}
Using Eqs. (\ref{eq:AniRatioMachOK}) for $r$ and $A_2$, we obtain
\begin{equation}
\beta_{\parallel 2} = \frac{1}{\sigma}\frac{\chi_1^2-1}{\chi_1^2}.
\end{equation}
The plasma is firehose stable if $A_2 \geq 1-1/\beta_{\parallel 2}$. This gives a first order equation for a critical $\sigma$ in terms of $\chi_1$, which solution is,
\begin{equation}\label{eq:sigmac1}
\sigma_{1c} = 1 -\frac{4}{\chi_1^2+3}-\frac{1}{\chi_1^2}.
\end{equation}

For a given $\chi_1$, the downstream after Stage 1 is stable for $\sigma \geq \sigma_{1c}$, and such a flow will not evolve beyond Stage 1. However, for $\sigma \
< \sigma_{1c}$, the magnetic field is not strong enough to stabilize the anisotropy generated in Stage 1. In this case, the plasma will undergo Stage 2, migrating towards the firehose stability threshold.

\section{Stage 2: Downstream on the firehose threshold}\label{sec:mod2}
Following our scenario, we now consider the case when the plasma after Stage 1 is unstable so that it evolves further towards Stage 2. We thus determine  the downstream properties of the plasma when its parallel and perpendicular temperatures lie on the firehose stability threshold.

\subsection{Density jump}
Here we  constrain the downstream temperatures imposing firehose stability. We therefore set,
\begin{equation}\label{eq:firehose2}
  \frac{T_{\perp 2}}{T_{\parallel 2}} = 1 - \frac{1}{\beta_{\parallel 2}}.
\end{equation}

As was done in section \ref{subsec:stag1} for Stage 1, we start from $U_2=(P_{2x}+n_2 k_BT_{\perp 2})/2n_2$. But we now use (\ref{eq:firehose2}) to express $T_{\perp 2}$ in terms of $T_{\parallel 2}$ and $\beta_{\parallel 2}$. From $\beta_{\parallel 2} = n_2k_BT_{\parallel 2}/B_0^8/8\pi$, we then obtain,
\begin{equation}
U_2 = \frac{3P_{2x} - 2B_0^2/8\pi}{2n_2}.
\end{equation}
Eq. (\ref{eq:conser3}) now gives for $P_{2x}$,
\begin{equation}\label{eq:Px22}
  P_{2x}= \frac{2}{5} n_2 \left[\frac{B_0^2/8\pi}{n_2}+  \frac{5}{2}\frac{P_1}{n_1}+V_1^2-V_2^2 \right].
\end{equation}
Equating with (\ref{eq:Px1}), which remains unchanged, we obtain an equation for $n_2$. Using the dimensionless variables defined in (\ref{eq:dimless}), we then obtain the following equation for the density ratio $r=n_2/n_1$,
\begin{equation}\label{eq:rs}
r^2 \left(\frac{5}{\chi_1^2} +1\right)-r \left(\frac{5}{\chi_1^2}+5-\sigma\right)+4 = 0.
\end{equation}
The solutions are,
\begin{eqnarray}\label{eq:ratioS}
  r_\pm &=& \frac{5+\chi_1^2 \left(5-\sigma \pm\sqrt{\Delta}\right)}{2 \left(\chi_1^2+5\right)},\\
 \Delta &=& \frac{25}{\chi_1^4}-\frac{10 (\sigma+3)}{\chi_1^2}+(\sigma-9) (\sigma-1).  \nonumber
\end{eqnarray}
Stage 2 does not allow solutions beyond a critical $\sigma$ where $\Delta < 0$\footnote{$\Delta$ turns positive again at large $\sigma$'s, but then the density jumps are negative.},
\begin{equation}\label{eq:sigmac2}
\sigma_{2c} =\frac{5}{\chi_1^2}+5-\frac{4 \sqrt{\chi_1^2+5}}{\chi_1}.
\end{equation}

\begin{figure}
  \begin{center}
   \includegraphics[width=.8\textwidth]{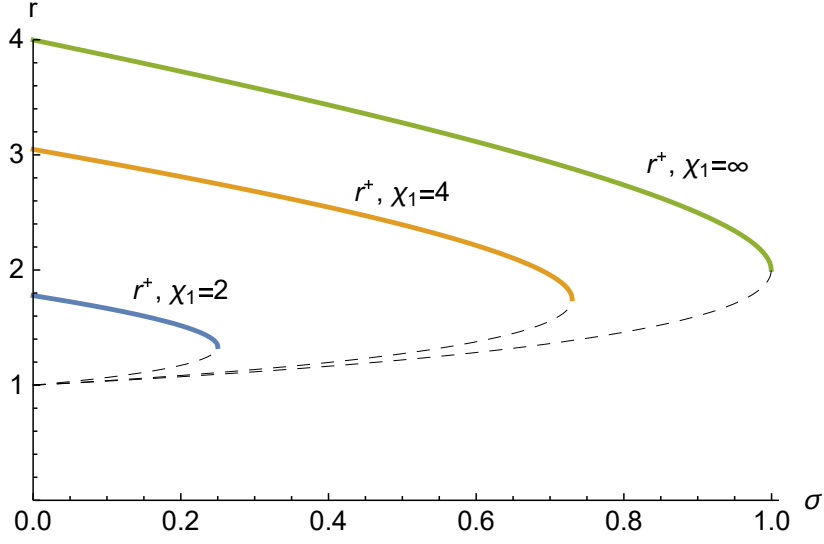}
  \end{center}
\caption{Density jump from Eq. (\ref{eq:ratioS})  as a function of $\sigma$ for various $\chi_1$. The $r^+$ branch is the shock solution, shown by solid lines. The dashed lines correspond to the $r^-$ branch. The model has no solution for $\sigma > \sigma_{2c}$ as given by Eq. (\ref{eq:sigmac2}).}
\label{fig:jmumpmod2}
\end{figure}

The density ratio defined by Eq. (\ref{eq:ratioS}) is plotted in Fig. \ref{fig:jmumpmod2}. For $\sigma=0$ and $\chi_1=\infty$, we have $r^+=4$ and $r^-=1$. These are the 2 solutions for the field-free problem in the strong shock limit for a 3D gas, so that we recognize that $r^+$ corresponds to the shock solution, while $r_-$ is the trivial solution. We will focus on the $r_+$ solution.

As an aside, note that for $\sigma \neq 0$, $r^-=1$ is no longer a solution. Why is this the case? In fluid mechanics, the solution $r=1$ also implies the absence of velocity jump and isotropic pressure jump. Such a solution is still retrieved in our analysis of Stage 1 because imposing  $T_\perp$ continuity is compatible with $r=1$. Yet, by imposing the firehose threshold (\ref{eq:firehose}) in Stage 2, we forbid the $r=1$ solution by rendering pressure continuity impossible, unless $\sigma=0$.

\subsection{Anisotropy}
We now come back to (\ref{eq:U2}), which gives
\begin{equation}\label{eq:U2Px}
  U_2  = \frac{P_{2x}}{2n_2}\left( 1+ 2 \frac{P_{\perp 2} }{P_{2x}} \right)  = \frac{P_{2x}}{2n_2} ( 1+ 2 A_2  ).
\end{equation}
Inserting this result into (\ref{eq:conser3}) gives the following equation for $A_2$,
\begin{equation}
  \frac{V_1^2}{2} + \frac{P_1}{n_1} + U_1  =  \frac{V_2^2}{2} + \frac{P_{2x}}{2n_2} ( 1+ 2 A_2  ) + \frac{P_{2x}}{n_2}.
\end{equation}
$P_{2x}$ is then eliminated through Eq. (\ref{eq:Px1}). Solving for $A_2$ and introducing the variables (\ref{eq:dimless}) gives,
\begin{equation}\label{eq:Aniso}
A_2  = \frac{1}{2}\frac{ \chi_1^2 (r-2) (r-1)+ r (5 r-3)}{\chi_1^2 (r-1)+ r}.
\end{equation}

\section{Physical interpretation of the parameter $\chi_1$}\label{sec:X}
When computing the Rankine-Hugoniot jump conditions in a fluid, the shock solution formally extends down to zero Mach number $\mathcal{M}$.
The reason why the range of physically allowed Mach numbers is restricted to $\mathcal{M} > 1$ is that the entropy jump is positive only over this range \citep{LandauFluid}.

In order to make physical sense of the parameter $\chi_1$, we therefore compute the entropy jump for both Stage 1 and Stage 2.

For a multi-temperature Maxwellian of the form,
\begin{equation}\label{eq:Max}
F= \frac{n}{\pi^{3/2} \sqrt{a} b}\exp \left(-\frac{v_x^2}{a}\right) \exp \left(-\frac{v_y^2+v_z^2}{b}\right),
\end{equation}
where $a=2k_BT_\parallel/m$ and $b=2k_BT_\perp/m$,
the entropy is
\begin{equation}\label{eq:entropy}
S = -k_B \int F \ln F = \frac{k_B}{2} n \left[ 3 + \ln (\pi^3 a b^2 )-2 \ln n\right].
\end{equation}
Computing the entropy-per-particle $s=S/n$ for the downstream and upstream plasma, we obtain the entropy jump across the shock,
\begin{eqnarray}\label{eq:entropydiff}
\frac{2}{k_B}\Delta s &=&  \frac{2}{k_B}(s_2-s_1) \nonumber \\
&=&  \ln \left( \frac{1}{A_2} \frac{ T_{\perp 2}^3}{ T_{\perp 1}^3}\right) - 2\ln r\,.
\end{eqnarray}

 In Stage 1, where there is no $T_\perp$ jump, we therefore have,
\begin{equation}
\frac{2}{k_B}\Delta s = -\ln (A_2 r^2).
\end{equation}

Using Eqs. (\ref{eq:AniRatioMachOK}) we find,
\begin{equation}
\frac{2}{k_B}\Delta s = \ln \left( \frac{( \chi_1^2-1) (3 + \chi_1^2)^3}{16 \chi_1^6}  \right).
\end{equation}

It is easily checked that $\partial \Delta s/\partial \chi_1 >0, \forall \chi_1 > 0$, and that $\Delta s=0$ for $\chi_1 = \sqrt{3}$. Hence, only $\chi_1 > \sqrt{3}$ is physically meaningful.
 We could therefore have defined a Mach number in Stage 1 through $\mathcal{M}_1^2=n_1V_1^2/3P_1$, consistent with an adiabatic index $\Gamma_{1D}=3$ for a 1D system.
  Indeed, since we freeze the perpendicular temperature, the system becomes effectively 1D. This can also be seen from Eq. (\ref{eq:AniRatioMachOK}), which tends to $r=2=(\Gamma_{1D}+1)/(\Gamma_{1D}-1)$ in the strong shock limit, and is larger than unity only for $\chi_1 > \sqrt{3}$.

For Stage 2, we start again from Eq. (\ref{eq:entropydiff}). Setting $T_{\perp 1} = T_1 =P_1/n_1 k_B$ and $T_{\perp 2} = A_2 T_{\parallel 2} = A_2 P_{2x}/n_2 k_B$, we obtain,
\begin{equation}
\frac{2}{k_B}\Delta s = \ln \left( \frac{r^5}{A_2^2} \frac{P_1^3}{P_{2x}^3} \right).
\end{equation}
The anisotropy $A_2$ is now given in terms of $r$ and $\chi_1$ by Eq. (\ref{eq:Aniso}). In turn, the density jump $r$ can be expressed in terms of $\chi_1$ and $\sigma$ through the $r_+$ branch of Eq. (\ref{eq:ratioS}). Finally, Eq. (\ref{eq:Px22}) allows us to express $P_{2x}$ in terms of $\chi_1$ and $\sigma$.

Through some manipulations, one can prove $\Delta s =0$ for $\chi_1=\sqrt{5/3}$ and $\sigma=0$. Stage 2 is therefore physically meaningful only for $\chi_1 > \sqrt{5/3}$, which is consistent with the definition of a 3D Mach number for this model, with $\mathcal{M}_1^2=n_1V_1^2/(5/3)P_1$. As a consequence, the density jump is not physical in Stage 2 for $\chi_1 < \sqrt{5/3}$, and one can check from Eq. (\ref{eq:sigmac2}) that $\sigma_{2c}(\chi_1 = \sqrt{5/3})=0$.

\bigskip

At this juncture, one might think that, for $\chi_1 \in [\sqrt{5/3},\sqrt{3}]$, our model predicts the downstream will simply settle down in Stage 2. However, this is not the case. The ansatz we made for the plasma imposes Stage 1 as the first stage of its downstream history. If $\chi_1 < \sqrt{3}$, there can be no Stage 1 shock, since such a shock would cause the entropy to decrease. Since there is no Stage 1 shock, there is no option for the gas to proceed further to Stage 2. Thus, the flow will be shock-free. In other words, our shock scenario makes physical sense only for $\chi_1 > \sqrt{3}$.

Could the plasma ``shortcut'' Stage 1, and jump directly to Stage 2
when $\chi_1 \in [\sqrt{5/3},\sqrt{3}]$? This is hard to say. In the
picture we have developed in this paper, the ``Stage 1 $\rightarrow$
Stage 2'' switch happens only if Stage 1 is  unstable. The plasma therefore comes to Stage 2, \emph{from an unstable
  position}. As the downstream plasma migrates towards stability, it
will therefore cross the firehose threshold coming from an unstable
position, and it settles down at the stability boundary. In contrast,
if Stage 1 cannot be attained because a 1D shock is unphysical
($\chi_1 < \sqrt{3}$, $\Delta s<0$), then the \emph{stable} downstream plasma has no reason to lie on the firehose stability threshold.  In that case, the
anisotropy cannot be related to the field in a deterministic way, for
the range of stable possibilities is infinite, as illustrated on
Fig. \ref{fig:firehose}.

Therefore, if we wish to define the ``usual'' Mach number, that is,
$\mathcal{M}_1^2=n_1V_1^2/(5/3)P_1 = (3/5)\chi_1^2$, our model offers
physical solutions only for $\mathcal{M}_1 > 3/\sqrt{5}\sim 1.34$.

Is this Mach number restriction physical, or simply an artifact of our model? As stated in the introduction, the fluid is disconnected from the field for a strictly parallel MHD shock. As a consequence, the minimum Mach number in the MHD regime is simply 1. Although to our knowledge no systematic study of weak collisionless shocks has been conducted so far, the $\mathcal{M}_1 > 3/\sqrt{5}$ restriction is probably an artifact of our model.

\begin{figure}
  \begin{center}
   \includegraphics[width=.5\textwidth]{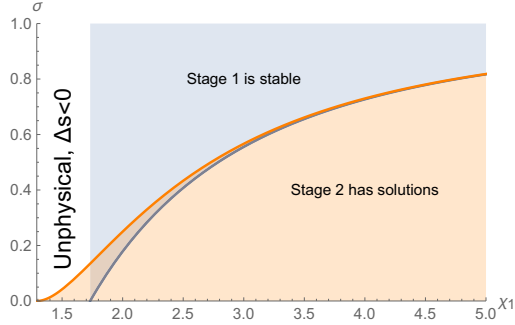}
  \end{center}
\caption{The downstream in Stage 1 is stable in the shaded blue area, and gives an entropy increase at the front only for $\chi_1>\sqrt{3}\sim 1.7$.  Stage 2 has solutions in the shaded orange region and gives an entropy increase at the front only for $\chi_1>\sqrt{5/3}\sim 1.3$ . Since our ansatz assumes the plasma goes first through Stage 1, our model is not physical for $\chi_1<\sqrt{3}$ (see discussion in section \ref{sec:X}). }
\label{fig:modelstogether}
\end{figure}

\section{Transition from Stage 1 to Stage 2}\label{sec:contact}
We have seen that Stage 2 has solutions only for $\sigma < \sigma_{2c}$ (Eq. \ref{eq:sigmac2}), while Stage 1 is stable only for $\sigma > \sigma_{1c}$ (Eq. \ref{eq:sigmac1}). Figure \ref{fig:modelstogether} shows the domains defined by these inequalities in the $(\chi_1,\sigma)$ plane.

Notably, the difference $(\sigma_{1c} - \sigma_{2c})$ goes to zero like $\chi_1^{-4}$ in the strong shock limit, so that both regions are exactly complimentary in this regime. We present a few results in this limit before we deal with intermediate values of $\chi_1$.

\subsection{Results for strong shocks, $\chi_1 \gg 1$}
Consider the case $\chi_1\rightarrow\infty$. For small values of $\sigma$, the point representing the system on Fig. \ref{fig:modelstogether} lies in the orange region. For larger $\sigma$'s, the system lies in the blue region. As can be seen on the figure, the blue and the orange regions do not overlap in the strong shock limit. Therefore, in this  limit, the system is either stable on Stage 1 (high $\sigma$'s), or firehose stable on Stage 2 because stage 1 was unstable (low $\sigma$'s).

Some useful analytical results can be derived in this regime.

Stage 1 is stable beyond $\sigma=\sigma_{1c} =1$. The density jump is $r_\infty=2$, and the anisotropy $A_{2\infty} \sim 4/\chi_1^2\to0$ as $\chi_1\to\infty$.

\begin{figure}
  \begin{center}
   \includegraphics[width=.9\textwidth]{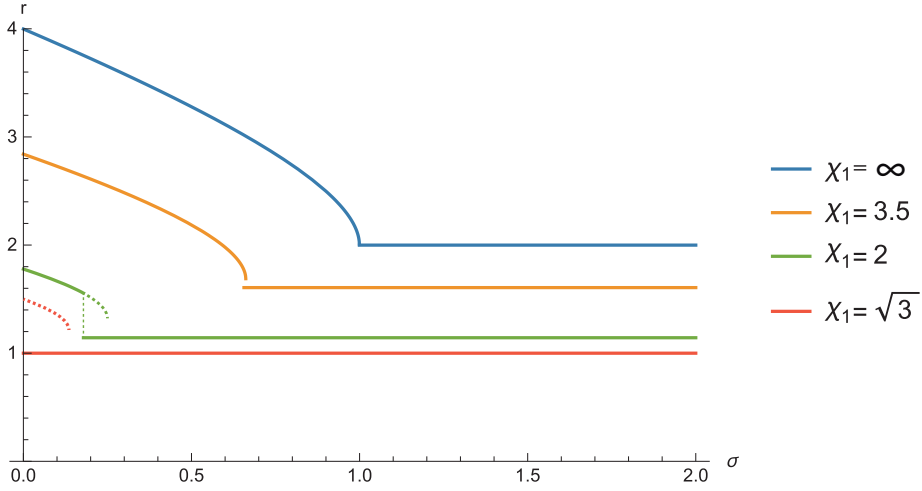}
  \end{center}
\caption{Density jump as a function of $\sigma$ for various values of $\chi_1$. Whenever Stage 1 (horizontal lines) and Stage 2 (curved lines) both have stable solutions, the physical one is Stage 1 (see section \ref{sec:interXi}). Stage 2 density jump is plotted for $\sigma \in [0, \sigma_{2c}]$. Stage 1 density jump is plotted for $\sigma \in [\sigma_{1c},2]$.}
\label{fig:jumpboth}
\end{figure}

For $\sigma<\sigma_{2c} =1$, Stage 2 has stable solutions. The density jump in Stage 2 reads,
\begin{equation}\label{eq:Rstrong}
r_\infty(\sigma)=\frac{1}{2} \left(\sqrt{(\sigma-9) (\sigma-1)}+5-\sigma\right),
\end{equation}
with $r_\infty(0)=4$ and $r_\infty(1)=2$ (see Fig. \ref{fig:jumpboth}). The anisotropy is given by
\begin{equation}\label{eq:Astrong}
A_{2\infty}(\sigma) = \frac{1}{4} \left(\sqrt{(\sigma-9) (\sigma-1)}+1-\sigma\right),
\end{equation}
with $A_{2\infty}(0)=1$ and $A_{2\infty}(\sigma\sim 1) \sim \sqrt{(1-\sigma)/2}\to 0$ as $\sigma\to 1$.

Eqs. (\ref{eq:Rstrong}, \ref{eq:Astrong}) clearly show that the anisotropy follows the same profile as the density ratio. Therefore, when $\sigma=0$, the strong shock is isotropic, with $A_2=1$. For small $\sigma$, the downstream anisotropy $A_2$ is given  Eq. (\ref{eq:Astrong}), which pertains to stage 2. For $\sigma=1$ and beyond, the system can remain in Stage 1, which is stabilized by the field. The anisotropy parameter is $A_2 =0$  (see eq.~\ref{eq:AniRatioMachOK} in the limit $\chi_1\to\infty$) all the way up to $\sigma = \infty$.

\subsection{Intermediate values of $\chi_1$}\label{sec:interXi}
For intermediate values of $\chi_1$, for example $\chi_1=2$, and small $\sigma$'s, Fig. \ref{fig:modelstogether} shows that Stage 1 is unstable. The system therefore goes to Stage 2, which offers solutions. Yet, for slightly larger values of $\sigma$, while Stage 1 is stable, at the same time Stage 2 also offers a solution. Which Stage should the system ``choose''? According to the plasma history we propose, the system goes first through Stage 1, before it switches to Stage 2, and \emph{the switch happens only if} Stage 1 is unstable. As a consequence, when stable, Stage 1 is always the physical solution. If not, the downstream plasma moves to Stage 2 which becomes its final state.

The density jump $r$ is eventually plotted in terms of $\sigma$ on Fig. \ref{fig:jumpboth}. The solution is given by Stage 2 at low $\sigma$, and by Stage 1 at high $\sigma$. When the function $r(\sigma)$ is multi-valued because Stage 1 is stable while Stage 2 offers solution, the physical solution is given by Stage 1. In such cases, we plotted the Stage 2 solution in dashed line. As evidenced, the departure from the MHD prediction (no-$\sigma$ dependance) increases with $\sigma$.

\begin{figure}
  \begin{center}
   \includegraphics[width=.9\textwidth]{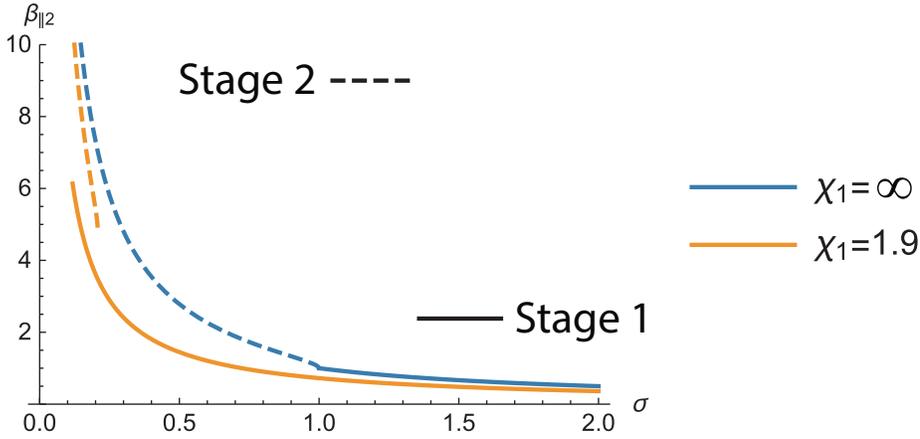}
  \end{center}
\caption{$\beta_{\parallel 2}$ as a function of $\sigma$ for Stage 1 (plain lines) and Stage 2 (dashed lines),  and $\chi_1=1.9$ and $\infty$. If $\sigma$ is such that it stabilizes Stage 1 while Stage 2 offers a solution, then the $\beta_{\parallel 2}$ of the downstream is given by Stage 1.}
\label{fig:beta(sigma)}
\end{figure}

\section{Locating the system on Figure \ref{fig:firehose}}\label{sec:betapara}
Although the description of the problem in terms of the parameters $(\sigma, \chi_1)$ is particularly adapted to PIC simulations, it is instructive to  locate the system on Fig. \ref{fig:firehose}, in terms of the parameters $(\beta_{\parallel 2}, T_{\perp 2}/T_{\parallel 2})$.

The anisotropy $A_2=T_{\perp 2}/T_{ \parallel 2}$ is given by Eq. (\ref{eq:AniRatioMachOK}) for Stage 1 and by Eq. (\ref{eq:Aniso}) for Stage 2. Regarding the parameter $\beta_{\parallel 2}$, it is given by Eq. (\ref{eq:BetaMod1}) for Stage 1. For Stage 2, it is readily derived from Eq. (\ref{eq:firehose}) in terms of the anisotropy (\ref{eq:Aniso}), since the firehose stability threshold is assumed for this stage.

Let us start by plotting $\beta_{\parallel 2}$ as a function of $\sigma$ for both Stages (Fig.~\ref{fig:beta(sigma)}). At large $\sigma$'s, Stage 1 is stable, so that the $\beta_{\parallel 2}$ of the downstream is given by this part of the plots (solid lines). For smaller values of $\sigma$, when Stage 1 becomes unstable, the $\beta_{\parallel 2}$ of the downstream is given by the Stage 2 part of the plots (dashed lines). If $\sigma$ is such that it stabilizes Stage 1 while Stage 2 also offers a solution, $\beta_{\parallel 2}$ is given by Stage 1.

\begin{figure}
  \begin{center}
   \includegraphics[width=.48\textwidth]{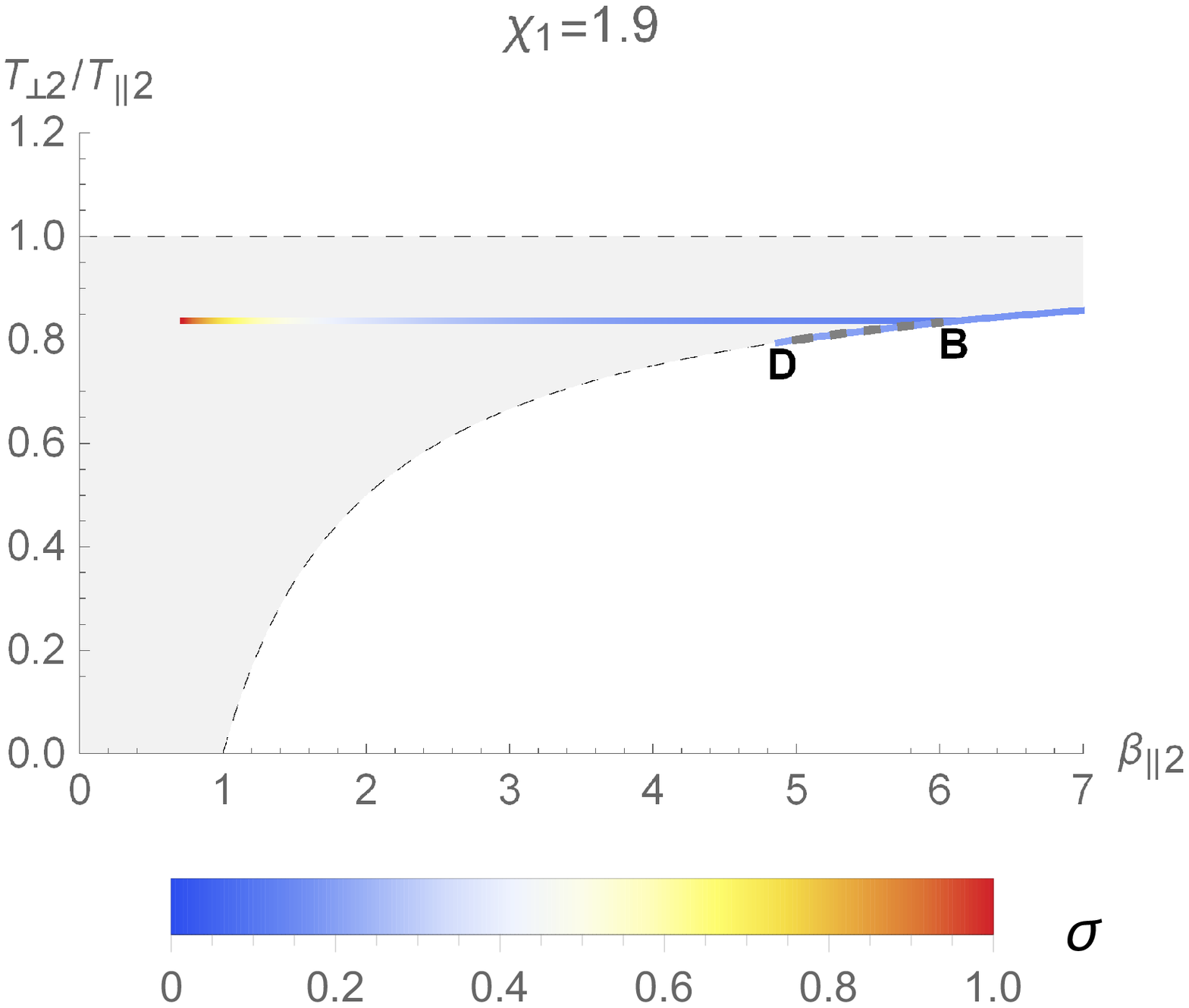}~~~\includegraphics[width=.48\textwidth]{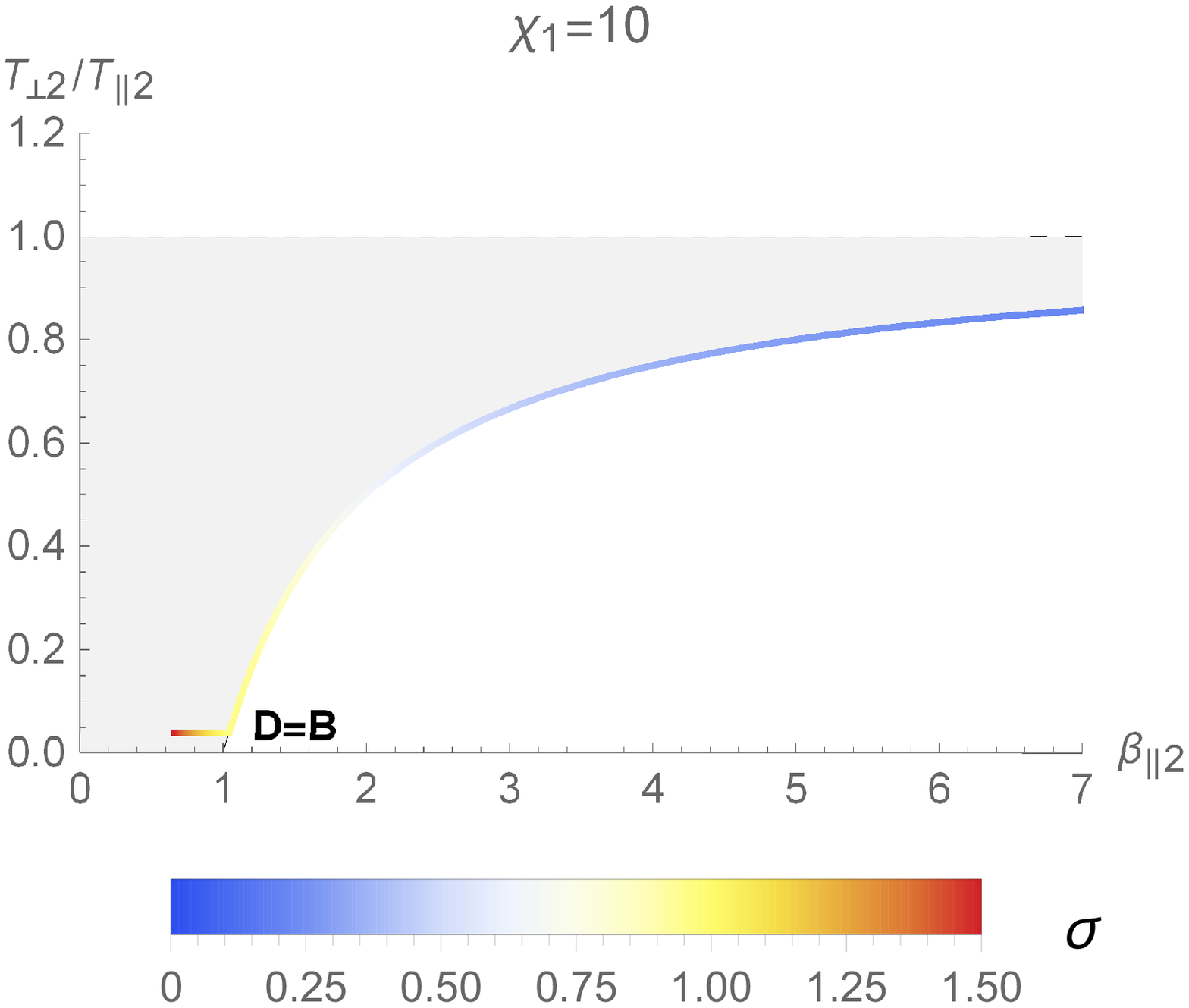}
  \end{center}
\caption{Location of the system on Fig. \ref{fig:firehose}, in terms of $\sigma$. For $\chi_1=1.9$, there is a range of $\sigma$'s where Stage 1 is stable while Stage 2 offers solutions. In this case, the system settles in Stage 1, namely, the first stage of its post-front history. For this reason, the portion \textbf{D-B} of the firehose curve is dashed since the system will not sweep it.}
\label{fig:path}
\end{figure}

We can now turn to Figure  \ref{fig:path} which locates the system on Fig. \ref{fig:firehose}. For both values of $\chi_1$ shown we plot the parametric curve,
\begin{eqnarray}
  X &=& \beta_{\parallel 2}(\sigma), \\
  Y &=& A_2(\sigma).
\end{eqnarray}
The resulting curves are colored in terms of $\sigma$. The point representing the downstream moves from right to left with increasing $\sigma$. For $\sigma \equiv \sigma_B$, the downstream is represented by point \textbf{B} of Figs. \ref{fig:firehose} \& \ref{fig:path}. For $\sigma  = \sigma_B + \varepsilon$, Stage 1 becomes stable while Stage 2 still has solutions. For Stage 2, the corresponding part of the curve is the short dashed \textbf{B}-\textbf{D} line in Fig. \ref{fig:firehose}, along the firehose threshold. This double solution is particulary visible on Fig. \ref{fig:path}-left, for $\chi_1=1.9$. For $\chi_1=10$, \textbf{B} and \textbf{D} are almost superimposed.

As explained previously, as soon as Stage 1 is stable, it is also the physical solution since our plasma first goes through this stage in the course of its downstream history. As a consequence, the system will not experience the \textbf{B}-\textbf{D} states of Stage 2 in Fig. \ref{fig:firehose}. In case $\sigma$ is in the corresponding range, the plasma will simply settle in the solution offered by Stage 1, namely, settle in the first stage of its post-front history.

\section{Conclusions}\label{sec:conclusions}
In order to derive an expression for the downstream temperature anisotropy in terms of the field strength in a parallel collisionless shock, we postulated a    particular history for the plasma. As it crosses the shock front, the plasma goes through the first stage of its history, namely ``Stage 1'', characterized by $T_{\perp 2}=T_{\perp 1}$. Stage 1 has $T_{\perp 2} /T_{\parallel 2} < 1$ and may or may not be firehose unstable. If the magnetic field is strong enough for Stage 1 to be stable, then the plasma remains in this state. However, if Stage 1 is firehose unstable,  the plasma moves to a new state located on the firehose stability threshold line. This is ``Stage 2''.

In both Stage 1 and Stage 2, the conservation equations permit a complete solution for the downstream properties of the plasma, including the temperature anisotropy parameter $A_2$. From this analysis, an effective adiabatic index $\Gamma$ can be computed from first principles. We start by writing the downstream internal energy as $U_2=\frac{1+2A_2}{2n_2}P_{x2}\equiv P_{x2}/(\Gamma-1)$, so that,
\begin{equation}\label{eq:Gamma}
  \Gamma = \frac{3+2A_2}{1+2A_2}
\end{equation}
fulfilling therefore $\Gamma(\sigma=0)=5/3$ and $\Gamma(\sigma>1)=3$.

The large-$\sigma$ plateau in the density jump observed in Fig. \ref{fig:jumpboth} is likely an artifact of our $T_\perp=$ constant hypothesis, rather than a real physical feature. For the strong shock case, for example, we do expect $\lim_{\sigma \to \infty} r = 2$, which the double adiabatic theory reproduces successfully. Yet, this theory turns exact only in the infinite $\sigma$ limit, where the Larmor radius is smaller than all the other length scales, shock front thickness included. But for moderate $\sigma$'s, the dissipation/heating occurring at the front should cause the plasma to deviate from $T_\perp=$ constant, even though the field may be strong enough to stabilize the first stage of the post-front evolution. The function $r(\sigma)$ is thus likely to be smoother than in our model.

This is indeed what was observed in \cite{BretJPP2017}. These simulations were relativistic, so that the non-relativistic theory described here cannot be directly compared to the numerical results. Forthcoming works should therefore focus on numerically testing the present theory.

As emphasized in the introduction, parallel shocks are excellent test beds to study departures from MHD, since according to MHD, the field and the fluid are disconnected. In such a configuration, any variation of shock properties with the field strength must be a kinetic effect. Yet, an effective adiabatic index like (\ref{eq:Gamma})  has to be obliquity-dependent if it is to be incorporated in MHD codes. Extending the present theory to oblique and perpendicular shocks is therefore necessary.

It should be possible to adapt the present treatment to perpendicular shocks by  considering the consequences of the double adiabatic theory for such systems. Instead of having $T_\perp$ conserved and $T_\parallel$ increased through the front, we would have $T_\perp \propto B$ increased by the density ratio\footnote{In a perpendicular shock, the field is amplified at the front by the same factor as the density \citep{Balogh2013}.}, and $T_\parallel \propto (n/B)^2$, constant. Stage 1 in the downstream will therefore have $T_\perp/T_\parallel > 1$, for which stability will be governed by the mirror instability \citep{Vogl2001}.

It would be useful to extend this work to the relativistic regime and the case of electron/ion plasmas, or even to account for the role of reflected particles at the front. Relativistic effects will affect the conservation equations, the stability thresholds and the double adiabatic invariants \citep{Scargle1968}. As already alluded to in the introduction, considering electron/ion plasmas may introduce temperature differences between species and extra physics downstream, whose roles are unclear at this stage. Finally, reflected particles may impact the conservation equations and the double adiabatic invariants.

\section{Acknowledgments}
A.B. acknowledges support by grants ENE2016-75703-R from the Spanish
Ministerio de Educaci\'{o}n and SBPLY/17/180501/000264 from the Junta de Comunidades de Castilla-La Mancha.  R.N. acknowledges support from the NSF
via grant AST-1816420. A.B. thanks the Black Hole Initiative (BHI) at
Harvard University for hospitality, and R.N. thanks the BHI for
support.  The BHI is funded by a grant from the John Templeton
Foundation.  The authors thank Reinhard Schlickeiser and Michael
Gedalin for valuable inputs, and the two referees for numerous helpful comments.


\begin{thebibliography}{26}
\expandafter\ifx\csname natexlab\endcsname\relax\def\natexlab#1{#1}\fi

\bibitem[Bale {\em et~al.\/}(2009)Bale, Kasper, Howes, Quataert, Salem \&
  Sundkvist]{BalePRL2009}
{\sc Bale, S.~D., Kasper, J.~C., Howes, G.~G., Quataert, E., Salem, C. \&
  Sundkvist, D.} 2009 Magnetic fluctuation power near proton temperature
  anisotropy instability thresholds in the solar wind. {\em Phys. Rev. Lett.\/}
  {\bf 103}, 211101.

\bibitem[Balogh \& Treumann(2013)]{Balogh2013}
{\sc Balogh, André \& Treumann, Rudolf~A} 2013 {\em {Physics of Collisionless
  Shocks: Space Plasma Shock Waves}\/}. New York: Springer.

\bibitem[Bittencourt(2013)]{bittencourt2013}
{\sc Bittencourt, J.A.} 2013 {\em Fundamentals of Plasma Physics\/}. Springer
  New York.

\bibitem[Bret(2016)]{BretJPP2016}
{\sc Bret, Antoine} 2016 Particle trajectories in weibel magnetic filaments
  with a flow-aligned magnetic field. {\em Journal of Plasma Physics\/} {\bf
  82}~(4), 905820403.

\bibitem[{Bret} {\em et~al.\/}(2017){Bret}, {Pe'er}, {Sironi}, {Sa{\c D}owski}
  \& {Narayan}]{BretJPP2017}
{\sc {Bret}, A., {Pe'er}, A., {Sironi}, L., {Sa{\c D}owski}, A. \& {Narayan},
  R.} 2017 {Kinetic inhibition of magnetohydrodynamics shocks in the vicinity
  of a parallel magnetic field}. {\em Journal of Plasma Physics\/} {\bf
  83}~(2), 715830201.

\bibitem[Chew {\em et~al.\/}(1956)Chew, Goldberger \& Low]{CGL1956}
{\sc Chew, G.~F., Goldberger, M.~L. \& Low, F.~E.} 1956 The boltzmann equation
  and the one-fluid hydromagnetic equations in the absence of particle
  collisions. {\em Proceedings of the Royal Society of London A: Mathematical,
  Physical and Engineering Sciences\/} {\bf 236}~(1204), 112--118.

\bibitem[{Erkaev} {\em et~al.\/}(2000){Erkaev}, {Vogl} \&
  {Biernat}]{Erkaev2000}
{\sc {Erkaev}, N.~V., {Vogl}, D.~F. \& {Biernat}, H.~K.} 2000 {Solution for
  jump conditions at fast shocks in an anisotropic magnetized plasma}. {\em
  Journal of Plasma Physics\/} {\bf 64}, 561--578.

\bibitem[Feynman {\em et~al.\/}(1963)Feynman, Leighton \& Sands]{FeynmanVol2}
{\sc Feynman, R.P., Leighton, R.B. \& Sands, M.L.} 1963 {\em The Feynman
  Lectures on Physics\/}. {\em The Feynman Lectures on Physics\/} v. 2.
  Pearson/Addison-Wesley.

\bibitem[{Gary}(1993)]{Gary1993}
{\sc {Gary}, S.~P.} 1993 {\em {Theory of Space Plasma Microinstabilities}\/}.

\bibitem[{Gary} \& {Karimabadi}(2009)]{Gary2009}
{\sc {Gary}, S.~P. \& {Karimabadi}, H.} 2009 {Fluctuations in electron-positron
  plasmas: Linear theory and implications for turbulence}. {\em Physics of
  Plasmas\/} {\bf 16}~(4), 042104.

\bibitem[Gerbig \& Schlickeiser(2011)]{Gerbig2011}
{\sc Gerbig, D. \& Schlickeiser, R.} 2011 Jump conditions for relativistic
  magnetohydrodynamic shocks in a gyrotropic plasma. {\em The Astrophysical
  Journal\/} {\bf 733}~(1), 32.

\bibitem[{Guo} {\em et~al.\/}(2017){Guo}, {Sironi} \& {Narayan}]{Guo2017}
{\sc {Guo}, X., {Sironi}, L. \& {Narayan}, R.} 2017 {Electron Heating in
  Low-Mach-number Perpendicular Shocks. I. Heating Mechanism}. {\em \apj\/}
  {\bf 851}, 134.

\bibitem[{Guo} {\em et~al.\/}(2018){Guo}, {Sironi} \& {Narayan}]{Guo2018}
{\sc {Guo}, X., {Sironi}, L. \& {Narayan}, R.} 2018 {Electron Heating in Low
  Mach Number Perpendicular Shocks. II. Dependence on the Pre-shock
  Conditions}. {\em \apj\/} {\bf 858}, 95.

\bibitem[Karimabadi {\em et~al.\/}(1995)Karimabadi, Krauss-Varban \&
  Omidi]{Karimabadi95}
{\sc Karimabadi, H., Krauss-Varban, D. \& Omidi, N.} 1995 Temperature
  anisotropy effects and the generation of anomalous slow shocks. {\em
  Geophysical Research Letters\/} {\bf 22}~(20), 2689--2692.

\bibitem[Kunz {\em et~al.\/}(2014)Kunz, Schekochihin \& Stone]{Kunz2014}
{\sc Kunz, Matthew~W., Schekochihin, Alexander~A. \& Stone, James~M.} 2014
  Firehose and mirror instabilities in a collisionless shearing plasma. {\em
  Phys. Rev. Lett.\/} {\bf 112}, 205003.

\bibitem[Landau \& Lifshitz(2013)]{LandauFluid}
{\sc Landau, L.D. \& Lifshitz, E.M.} 2013 {\em Fluid Mechanics\/}. Elsevier
  Science.

\bibitem[Lichnerowicz(1976)]{Lichnerowicz1976}
{\sc Lichnerowicz, Andr\'{e}} 1976 Shock waves in relativistic
  magnetohydrodynamics under general assumptions. {\em Journal of Mathematical
  Physics\/} {\bf 17}~(12), 2135--2142.

\bibitem[{Majorana} \& {Anile}(1987)]{Majorana1987}
{\sc {Majorana}, A. \& {Anile}, A.~M.} 1987 {Magnetoacoustic shock waves in a
  relativistic gas}. {\em Physics of Fluids\/} {\bf 30}, 3045--3049.

\bibitem[Marcowith {\em et~al.\/}(2016)Marcowith, Bret, Bykov, Dieckmann,
  Drury, Lemb\`{e}ge, Lemoine, Morlino, Murphy, Pelletier, Plotnikov, Reville,
  Riquelme, Sironi \& {Stockem-Novo}]{Marco2016}
{\sc Marcowith, A, Bret, A, Bykov, A, Dieckmann, M~E, Drury, LO'C,
  Lemb\`{e}ge, B, Lemoine, M, Morlino, G, Murphy, G, Pelletier, G, Plotnikov,
  I, Reville, B, Riquelme, M, Sironi, L \& {Stockem-Novo}, A} 2016 The
  microphysics of collisionless shock waves. {\em Reports on Progress in
  Physics\/} {\bf 79}, 046901.

\bibitem[Maruca {\em et~al.\/}(2011)Maruca, Kasper \& Bale]{MarucaPRL2011}
{\sc Maruca, B.~A., Kasper, J.~C. \& Bale, S.~D.} 2011 What are the relative
  roles of heating and cooling in generating solar wind temperature
  anisotropies? {\em Phys. Rev. Lett.\/} {\bf 107}, 201101.

\bibitem[{Scargle}(1968)]{Scargle1968}
{\sc {Scargle}, J.~D.} 1968 {On Relativistic Magnetohydrodynamics}. {\em
  \apj\/} {\bf 151}, 791.

\bibitem[{Schlickeiser}(2010)]{Schlickeiser2010}
{\sc {Schlickeiser}, R.} 2010 {Linear Theory of Temperature Anisotropy
  Instabilities in Magnetized Thermal Pair Plasmas}. {\em The Open Plasma
  Physics Journal\/} {\bf 3}, 1--19.

\bibitem[Schlickeiser {\em et~al.\/}(2011)Schlickeiser, Michno, Ibscher, Lazar
  \& Skoda]{SchlickeiserPRL2011}
{\sc Schlickeiser, R., Michno, M.~J., Ibscher, D., Lazar, M. \& Skoda, T.} 2011
  Modified temperature-anisotropy instability thresholds in the solar wind.
  {\em Phys. Rev. Lett.\/} {\bf 107}, 201102.

\bibitem[{Sironi} \& {Spitkovsky}(2011)]{Sironi2011ApJ}
{\sc {Sironi}, L. \& {Spitkovsky}, A.} 2011 {Particle Acceleration in
  Relativistic Magnetized Collisionless Electron-Ion Shocks}. {\em \apj\/} {\bf
  726}, 75.

\bibitem[Vogl {\em et~al.\/}(2001)Vogl, Biernat, Erkaev, Farrugia \&
  M\"uhlbachler]{Vogl2001}
{\sc Vogl, D.~F., Biernat, H.~K., Erkaev, N.~V., Farrugia, C.~J. \&
  M\"uhlbachler, S.} 2001 Jump conditions for pressure anisotropy and
  comparison with the earth's bow shock. {\em Nonlinear Processes in
  Geophysics\/} {\bf 8}~(3), 167--174.

\bibitem[Weibel(1959)]{Weibel}
{\sc Weibel, E.~S.} 1959 Spontaneously growing transverse waves in a plasma due
  to an anisotropic velocity distribution. {\em Phys. Rev. Lett.\/} {\bf 2},
  83.

\end{thebibliography}

\end{document}